# Group-centered framework towards a positive design of digital collaboration in global settings

Irawan Nurhas[1], Jan Pawlowski[2], Stefan Geisler[1] and Maria Kovtunenko[2]

Institute of Positive Computing[1], Institute of Computer Science[2], Ruhr West University of Applied Sciences
Germany
{irawan.nurhas, jan.pawlowski, stefan.geisler, maria.kovtunenko}@hs-ruhrwest.de

Bayu Rima Aditya
School of Applied Sciences
Telkom University
Indonesia
bayu@tass.telkomuniversity.ac.id

*Abstract*— globally distributed groups require collaborative systems to support their work. Besides being able to support the teamwork, these systems also should promote well-being and maximize the human potential that leads to an engaging system and joyful experience. Designing such system is a significant challenge and requires a thorough understanding of group work. We used the field theory as a lens to view the essential aspects of group motivation and then utilized collaboration personas to analyze the elements of group work. We integrated well-being determinants as engagement factors to develop a group-centered framework for digital collaboration in a global setting. Based on the outcomes, we proposed a conceptual framework to design an engaging collaborative system and recommend system values that can be used to evaluate the system further.

## I. INTRODUCTION

Information technology (IT) provides excellent possibilities for global collaboration. At the same time, online collaboration presents barriers and challenges that can hinder the willingness to collaborate, to complete tasks and generally to use the system for group work. It is important to tackle these barriers and encourage users to participate, but this requires extra effort and interdisciplinary collaboration to ensure that the system is useful for the user and the business. To maximize the potential of IT for collaborative work, the system must be designed based on user requirements. Therefore, a preliminary user study is required before the system is created. Cooper et. al. [1] recommend using persona as a tool to model the user that contains information about behaviors, goals, and other factors. Persona also can be used as an interdisciplinary communication tool within and outside an organization. As personas contain user information, they can be crucial for the whole development process, from prototyping to system evaluation and for both product design and marketing strategy.

Developers should not only design collaborative systems but also deliver a joyful experience that encourages the user to complete the collaborative task. Therefore, the term of "positive computing" is applied by Calvo et. al. [5] to design systems that are not only useful but also can promote human well-being (including happiness, human potential, joyful experiences and task accomplishment). Furthermore, Pawlowski et. al. [6] explain how the "trend" toward positive computing could influence the design process for business information systems. In the positive computing, technology is designed to improve the user's well-being, one valuable recommendation to the design process is by integrating well-being determinants into the early phase of design [3]. The principles of positive computing can be implemented in many technological areas, including wearable technologies, interactive communication tools, mobile apps and technologies in the work environment [7]. To our knowledge, there is a lack of group-based design guidelines for positive design of a collaboration system that shifts the designer's focus from an individual view to a group view. Under those circumstances, in this study, we conducted a qualitative design science research [8] that aimed to address the following research questions:

1) What are the group aspects that need to be addressed in the design of an engagement collaboration system?
2) How can a system design for an engaging collaboration be modeled and applied to the design process?

By answering these research questions, this study contributes to provide a conceptual model for an engaging collaborative system based on the applicable tool in the development process of system design (persona). Next, we discuss the theoretical background of our model.

## II. STATE OF THE ART

Cabrero et al. [28] and Matthews et al. [2] provide an overview of group attributes that are important in personas for designing collaborative software (i.e., group goal, working group style, collaboration needs, members, and roles) across culture [28]. However, both [28] and collaboration-personas' (CP) [2] attributes as well as international personas [29] do not provide information related to the virtue and pleasure for positive design [3] and it is important to realize that the root issue of positive design is to promote well-being. Correspondingly, the PERMA (Positive emotion, Engagement, Relatedness, Meaning, and Accomplishment) well-being model was introduced as a new understanding of well-being and happiness [4] that combines hedonic and eudaimonic approaches as well-being factors [5]. Even though the PERMA model does not provide a holistic picture of the digital collaboration system, it contains elements (e.g. positive emotions, engagement and relatedness or empathy) that can improve overall well-being and also can be integrated into system design [5].

In this study, we used the field theory or Lewin's field theory [9] [10] [11], which is described as (1) in social psychology.

$$B = f(p,e) \quad (1)$$

Through (1), Lewin shows that the behavior ($B$) is influenced by the interactions of a person ($p$) with the environment ($e$). The equation shows that p and e are not independent of one another and can be mutually influential [9][10]. On the one hand, the behavior ($B$) occurs because people interact with their environments (extrinsic motivation). On the other hand, people's behavior also is influenced by their intrinsic motivation [12]. Both interactions can be expressed and explored through acts and through a person's dynamic field, thoughts, and emotions [9][10][13].

The field theory was adopted because it describes the important factors that affect the individual behaviors within an organization. In field theory, the focus of the environment is person-to-person or multi-person interaction [9], which is relevant for group work. Based on Lewin's field theory we concluded three important relational elements (actions, emotion, and interaction) [9][10][13][14] to design a joyful experience. These elements can be seen in Figure 1.

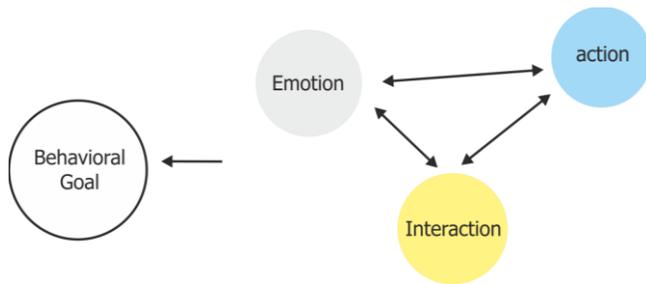

Fig. 1. Derivation of Lewin's field theory

Figure 1 was used to describe the behavioral goal; the expected behavioral group goal can be achieved through the design of an environmental condition that contains those elements. In the following section, we present the conceptual model for designing an engaging collaborative system.

## III. METHOD

The study applied the design science research methodology [8]. Peffers et al. [8] developed this method by adopting Design Science Research (DSR) into the field of information systems. DSR is prepared to focus on problem solutions by providing artifacts in the form of models, tools, prototypes or end working systems. Based on DSR [8] several steps are carried out in this study: identification of the solution to develop the conceptual model through a structured literature review and analysis of the data based on the theoretical background. Next, demonstration and evaluation of the proposed model through interview and case study. Finally, the result will be published to the scientific community. The results of the individual steps are explained in more detail in the next section.

## IV. DEVELOPMENT OF A CONCEPTUAL FRAMEWORK

The development of the concept consisted of several stages; we started our research process with a structured literature review. Then, we analyzed the information we had gathered to develop our model. We designed our model based on analyzing the persona attributes, well-being determinants and requirements of global system. The steps of conceptual development are explained in greater detail below:

1) Sample and data collection

The first stage of model development was to collect existing research on persona through structured literature reviews [15]. We read each abstract carefully to filter inappropriate articles. Then, we analyzed 89 articles focused on persona attributes at the individual and group level.

2) Analysis of the aspects to integrate well-being determinants into persona

Next, we grouped and analyzed the personas attributes to identify the integration aspects of well-being and personas. The integration aspects are grouped into technical, social and motivational, instrumental, operational and adaptational aspects as important aspects for the integration of well-being elements into persona attributes [16].

3) Analysis of the disadvantages and advantages of collaboration-personas

The selection of CP as basic attributes was based on two main criteria: adaptability for individual personas and understandability (Matthews et. al. [2] gives important group persona attributes accompanied by a case study and discussion of the use of each attribute). The selection of these criteria clarified how the attributes can be used and developed further.

In the field theory, there are two elements in addition to action that influence the particular behavior: the interaction and the resulting emotion. (In this case, we defined the interaction as user-to-user interaction through a digital system) None of these aspects are mentioned in [2]. The attributes in [2] focus primarily on information related to the group actions (that is, the group goal as it relates to collaboration activities, objectives of the collaboration, and the type of activity that occurs through collaboration). Furthermore, there is a lack of information on how these issues can be seen from a global point of view.

4) Analysis of relation between collaboration-personas with positive-personas aspects

After we evaluated the attributes of personas related to the collaboration (stage 3) and identified the aspects which require the integration of well-being determinants into personas (stage 2), we used PERMA well-being model. The PERMA was used because the model does not only include emotional (positive emotion) and social interaction (relatedness) but also an action-aspect (accomplishment), which is necessary to analyze the attributes derived in stage 3 and their relation to the aspects identified in stage 2. The important element of CP and their relationship with the integrating aspects in the PERMA model are explained below:

a) Group goal and instrumental-aspects

The group goal is the reason why people work together on a project. They act and react collectively based on emotional situations. Therefore, a **collective emotional goal** [17][18] can be used to engage a group of people in acting and collaborating on the same project. Positive emotions that provide information for the system designer include joy, interest, contentment and love [19]

b) Collaboration needs and motivational- and social-aspects

The group goal provides information about the collective emotional goal, whereas collaboration needs are social and individual reasons to cooperate in group work, or motivational background that related to the interaction or feeling with other people (e.g., empathy) [5][4]. Examples of design elements are feedback, response and recognition, which are **socially meaningful objectives** [20] that can be facilitated by the system to encourage collaboration.

c) Group working style and timing-aspects

In exploring CP, it is important to identify the working group style [2], which describes the different types of activities a diverse group can engage in to achieve the collective goals and objectives. To design an engagement system, the researcher must identify the place and structure of these activities properly [23], so that the user can observe and feel the flowable experiences in the process of collaboration. Therefore, it is important to identify **the observable collective activities** [21][22]. The type of working styles that form the collaboration styles are dynamic, stable, community, committee, client–supplier, and professional [2]. The important observable activity-related attributes that can improve well-being are **dosage, frequency, timing, variety, and trigger/social trigger** [23].

d) Members, roles and operational- and adaptive-aspects

The group is a body of people who take on different responsibilities. Through shared responsibilities, the group can combine tasks to achieve the goal. **Accomplishment** of a job through a **measurable task** [24] does not only support group work but also triggers feelings of self-esteem and self-compassion that lead to feelings of belonging and integration with the team, including the design that can accommodate the physical and cognitive user condition.

The analysis of the relation between the CP and the positive-personas aspects provide an overview of how well-being elements are useful and important in the group to encourage collaboration (i.e., **the contextual aspects**).

5) Analysis of global issues in a collaborative system

Next, we analyzed the effect of the global ecosystem on the design of a digital collaboration system. Based on our development of a collaborative system [25], two broad categories emerged as the primary requirements for a collaborative system at the global level. The first requirement category to access the system from different locations is the basic need for an online distributed system, which refers to **accessibility of technology**. The second requirement is a system that can accommodate the difference in working time zone and language as well as the perception of colors, layouts, and symbols. This second requirement refers to **intercultural adaptability**. These two requirements are closely related to user-system interaction (i.e., **the technical aspects**).

V. RESULTS AND DISCUSSION

Based on the elements in Figure.1 and stages of analysis in section three, we defined two important aspects to improve user involvement in the collaborative system. The first aspect is the technical and represents the interaction elements between the user and the system. The technical aspect is the primary feature of any collaborative system intended for a global ecosystem. Before designing a system to encourage collaboration in a global ecosystem, the designer must be sure to address this technical aspect.

The next aspect is contextual and represents the emotions and actions of the group. The contextual aspect defines the elements of well-being and the integration of these elements to design a system for engaging user in collaboration. The designer can develop the technical aspects that support user-system interaction by analyzing the information from a contextual point of view. Our conceptual framework was based on the persona attributes, and a development team could use our framework to model the user in the form of a persona as the basis for designing and evaluating a system. Figure 2 presents an overview of our conceptual framework.

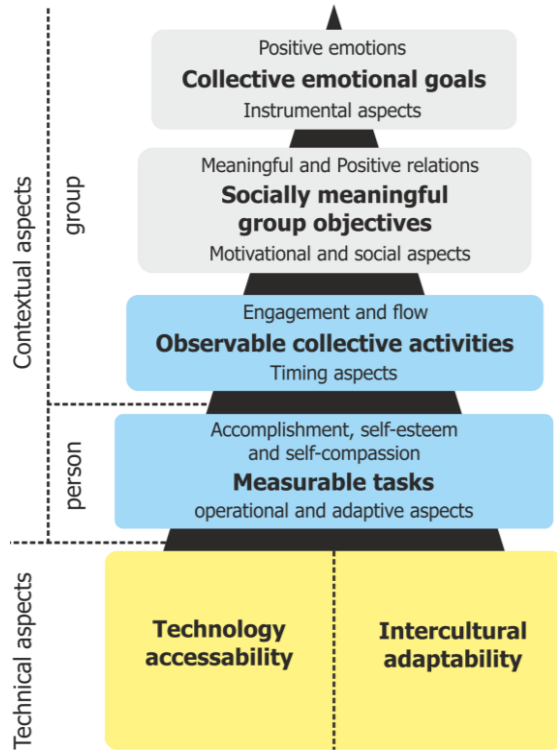

Fig. 2. Collaboration-positive-personas.

Our framework shows the adaptation of CP with well-being determinants. Because the framework is based on personas, it can be used to design a persona template and attributes. Therefore, we will provide a template to model a group for preliminary study of system design (still in progress).

Accordingly, we present our value-based system overview for a digital collaborative system. Figure 3 shows the proposed values for a collaborative system in a global environment. The value-based overview was developed by identifying the system characteristic related to important elements in Figure 2. The system values based on our model are as follows:

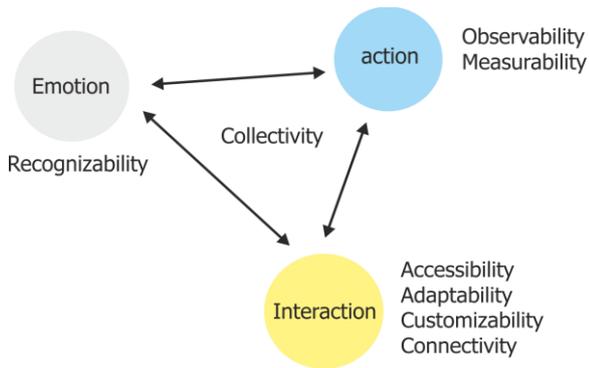

Fig. 3. Value-based system overview

- Collectivity

This value is related to all aspects in the Figure 2. The collective emotional goals, social objectives, activities, and tasks are important aspects to recognize a collaboration in a group work.

- Recognizability

This value is related to the positive emotional goal. The clarity of the emotional goal can help to determine the design elements or the form of design interaction between the user group and the system.

- Measurability

Action-measurability is related to the accomplishment of the group task (operational and adaptational aspects). The individual capability (competence required to collaborate in the group, physical and cognitive condition) is important to recognize, that can help the user to complete the task. Moreover, a measured task that can easily be completed improves the user's feeling of confidence in the early stage of system adaptation.

- Observability

Action-observability is related to the proper visible and structured process and moment-to-moment activity. This value includes the dosage, variety, and timing of a collaboration activity that can be observed in collective way (e.g., communication, coordination, and task management).

- Accessability

This feature relates to the technical aspects of the user-system interaction. To support online collaborative systems, it is important that the system can be accessed from diverse locations through the technology availability of different users and the abilities of each user.

- Adaptability

This value is related to the technical aspects of the digital system which facilitates physical and cognitive aspects of the user. The design of the system should adapt to the physical and cognitive conditions of each of the users in the group.

TABLE I. SAMPLE OF USE

| Collective positive emotion |
|---|
| Interest or feeling to become involved, having new experiences with a person or object and the feeling characterized by openness to new ideas.<br>Design implications:<br>● A system that provides a list of shared idea and a short detail information about the idea.<br>● Providing information about free license and the benefit of using the open educational resources.<br>● Providing suggestions about an idea that is maybe relevant to the skills and interest of the user. |
| Socially meaningful objectives |
| ● Provide open knowledge for other people<br>● Knowledge exchange<br>Design implication:<br>● Provide information about impact of the developed material for others (for example number of download, downloads location, review from other) |
| Observable activities |
| Development of content structure<br>Design implications:<br>● Real-time concurrent editing feature to edit the content and to show what others are doing and what is done.<br>● Ability to show historical changes of the content.<br>● Provide asynchronous communication (e.g. messaging).<br>● Provide reminders/notifications that appear when something changes in the content structure. |
| Measurably accomplished tasks |
| Complete the content of materials<br>Design implications:<br>● Provide list of completed jobs.<br>● Provide digital badges of task accomplishment. |
| Technology accessibility |
| Design implications:<br>● The use of web-based technology by utilizing HTML5 that can be freely assessed through different platforms. And it is accessible for users in developed and developing countries.<br>● Requires no installation to use the system. |
| Intercultural adaptability |
| Design implication:<br>● Modularity of system setting that can be easily added and customized through the developer for different cultures. |

- Customizability

This value is related to the technical aspects of the global settings. The customizability is the ability of the system to address the different values in the separate locations of group members as well as individual needs and preferences.

- Connectivity

Connectivity is a technical aspect of the system that can help to improve the relatedness among members through diverse types of communication and coordination channels.

Next, we provide an example of how the model can be used to design a collaborative system in global setting. The purpose of this example was to understand the implementation of the model. We used the scenario described in [25] to design an authoring system for collaborative creation of OER (Open Educational Resources) within the global scope. The following Table 1 describes the various elements of our model.

The proposed value framework is slightly different from the engagement framework presented in [26] that is a general framework developed for a gamification [26]. Our model shares the shame aspects (meaning, accomplishment, and relatedness) but differs in its inclusion of the collectivity of emotional goals, observable activity, intercultural adaptability and technological accessibility, which are our aspect of group work in global setting. Therefore, we consider it necessary to discuss the model with experts or practitioners and to evaluate the framework through qualitative interviews.

## VI. EVALUATION

We interviewed eight researchers and practitioners with a computer science and business information background (Gender: 6 males and 2 females; Age: $M = 31$; Years of experience: 3-9 years ($M = 6.25$)), including three participants from Germany and five from Indonesia, to discuss and evaluate our model. We used evaluation criteria provided by Frank [27] and we evaluated the proposed model with regard to benefit, knowledge transfers as well as deployment perspectives. Some of the important notes from the discussion are presented below:

1) A guideline for questions will be helpful, for the implementation and the integration of the model (the contextual aspects) into the persona tool.

2) The framework can accommodate different views of an interdisciplinary team by providing pre-information about the impacts of well-being and the corresponding shared well-being perspective of the user.

3) The model provides important aspects of an engaging system in global collaboration by accommodating cultural barriers and differences in technology availability. However, detail explanation is still required due to unfamiliarity with the graphical representation of the model. Moreover, an example how the framework can be integrated into collaborative software development process (e.g., agile).

Despite the fact that the results of the evaluation were in the right direction, our model still must be validated in a different context through more perspectives from other experts and practitioners. Therefore, it is important to communicate our results with the larger research community to gain more valuable feedback.

## VII. CONCLUSION AND FUTURE WORK

Our framework is a group-focused design that demonstrates contextual and technical aspects of a collaborative system. The practical contribution of the model should serve as a guide for collecting useful data to implement the positive design in group work. This study contributes to the theoretical aspect of the design process by adding elements to the existing persona [2] which integrate well-being determinants of collaborative work to improve user engagement in the global collaboration. Additionally, the study contributes to provide evaluation criteria based on derived values. However, the framework is not intended to eliminate the individual user approaches.

ACKNOWLEDGMENT

We are grateful for the support from institute of positive computing of Ruhr West University of Applied Sciences which was funded by the ministry for Innovation, Science, Research and Technology of the State of North Rhine-Westphalia.